# Physics at the Munich Tandem Accelerator Laboratory

Günther Dollinger
Universität der Bundeswehr, München, and
Excellence Cluster „Munich-Centre for Advanced Photonics"

Thomas Faestermann
Physik Department, Technische Universität München, and
Excellence Cluster „Origin and Structure of the Universe"

## Abstract

This review reports on the science performed in various fields at the Munich tandem accelerator during the past decade. It covers nuclear structure studies, also with respect to astro- and particle physics as well as for the understanding of fundamental symmetries, the extremely sensitive detection of long-lived radionuclides from Supernova or r-process production with accelerator mass spectrometry and studies of the elemental composition of thin films with extreme depth resolution and sensitivity by elastic recoil detection (ERD). The ion microbeam is used for 3D hydrogen microscopy as well as in radiobiology to study the response of living cells on well-defined irradiations. In medical research new therapeutic methods of tumour irradiation are tested using proton minibeams as well as the determination of ion ranges in tissue with iono-acoustics. Primary and secondary beams from the accelerator are also used for development and testing of detector components in large setups, e.g. at the LHC, and for testing new kinds of fuel materials of high uranium density to use them as medium enriched fuels at the Munich research reactor FRM II in the future.

## The MP Tandem Accelerator

The Tandem accelerator installed at the Accelerator Laboratory in Garching, just 20 km north of Munich, is of the "Emperor" (MP) series manufactured by High Voltage Engineering corporation (HVEC). It delivered the first beams for experiments in 1970 and came close to its design voltage of 10 MV [Ass74].

In 1975 a phase of improvements to the accelerator started. A Pelletron charging system was installed and also acceleration tubes from NEC and a Portico [Mue84]. But finally in 1991 the tubes were exchanged again to the extended version of HVEC with very good results [Mue93]. Routine operation at a terminal voltage of 14 MV was then possible. In 1996 we installed the 90° analysing magnet which we had obtained from the Daresbury laboratory [Car97]. It has a bending radius of 1.667 m, calibrated with the spectrograph [Fae97] and a maximum field of 1.6 T. Thus it is powerful enough to even analyze ions of very heavy nuclei in their most probable charge states at a terminal voltage of 10 MV.

As negative ion sources we use a single-cathode sputter source for most ions with moderate intensity. For very intense hydrogen-beams of high brilliance there is a commercial multicusp ion source [Mos12]. Polarized hydrogen-beams and also helium-beams can be produced by an atomic-beam ion source [Her05] with an ECR-ionizer and a subsequent charge-exchange cell for double electron capture in a Cs-vapour. The polarization is obtained by a Stern-Gerlach separation and RF-transitions for the transfer of the electronic polarization to the nuclei. In addition there is, for AMS experiments, a dedicated port for sputter sources connected to a 90° analyzing magnet that provides a mass resolution M/δM ≈ 200-300. A pulsed beam with a width of about 2 ns can be produced with a chopper and a buncher on the low energy side and a chopper behind the tandem.

In 2002 the Maier-Leibnitz-Laboratory (MLL) was founded by the Ludwigs-Maximilians-Universität and the Technische Universität München, which now has the accelerator lab as one of its research activities besides high-energy, neutron, laser and medical physics.

In a "laboratory portrait" 17 years ago [Fae01] a report on the lab was given with a special emphasis on interdisciplinary research. Here we also want to emphasize the important role, that our Q3D spectrograph played in the last decade in nuclear structure physics and its impetus on neighbouring fundamental disciplines, as astro- and particle physics.

## The Q3D Magnetic Spectrograph

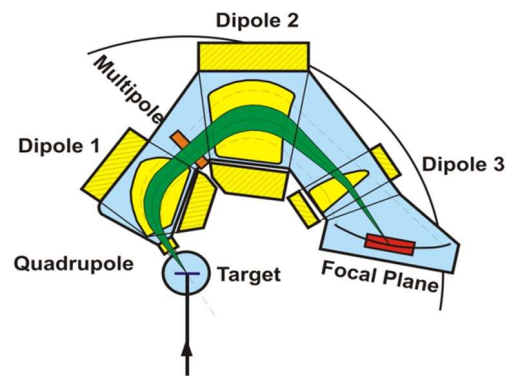

**Figure 1.** *Sketch of the Q3D spectrograph.*

Today, the Q3D magnetic spectrograph at the MLL is world-wide unique as a precision instrument for nuclear structure physics. It was designed as a double-focusing instrument by Harald Enge [Eng70] from MIT around 1970 after he had designed the split-pole spectrograph and several others. The first two were built identically for Heidelberg and Munich (Fig. 1). They were the first intended for detectors with electronic readout and therefore had a huge dispersion of $\delta x/(\delta p/p) \approx 20$ cm/% along the focal plane, which is actually bent and tilted by about 45° against the incoming particles. The bending radius is 0.95 - 1.05 $m$ and the maximum field about 1.6 T. The Munich version later on obtained a stronger quadrupole with independent power supply. The accepted solid angle is 14 msr, with a horizontal angle of 6.8°. For the correction of the kinematic walk over this angular range a magnetic multipole element was provided between 1$^{st}$ and 2$^{nd}$ dipole, where there is an intermediate vertical focus. Using this an energy resolution of $2\times10^{-4}$ (FWHM) over about 1.5 m of the focal plane can be achieved. Angular distributions can be measured between 0° and about 140°. As detectors for light particles (p,d,t,$^3$He,$\alpha$) in the focal plane we use most often a 0.9 m long detector, based on two single-wire proportional counters, one with a readout of 3 mm wide cathode strips for position ($\delta x \sim 0.1$ mm) and energy-loss information and a scintillator for the residual energy [Wir00]. For heavy ions we have two detectors: both use proportional counters for position and energy loss, one measures the residual energy in a 1 $m$ long ionization chamber [May80], the other one in 128 Si-detectors, each 11 $mm$ wide [Alb94].

## Nuclear Structure Studies

### The heaviest doubly magic nucleus $^{208}$Pb

Excited states in $^{208}$Pb have been studied extensively at the Q3D in the last decade. All the information gathered so far has been summarized in a detailed paper [Heu16]. Besides $^{208}$Pb(d,d') and $^{206,207,208}$Pb(d,p) reactions a special type of $^{208}$Pb(p,p') reaction has been used, where the energy of the projectile is chosen such that it populates states in $^{209}$Bi which are isobaric analog states (IAS) of $^{209}$Pb states. The outgoing proton then leaves the final nucleus in particle-hole states, where the particle is determined by the IAS. I.e. this type of reaction is equivalent to a neutron pickup from a single particle state in $^{209}$Pb. Thus, neutron particle-hole states have been populated with the particle in the orbitals $g_{9/2}$, $i_{11/2}$, $j_{15/2}$, $d_{5/2}$, $s_{1/2}$, $g_{7/2}$, $d_{3/2}$. These resonances are reached with proton energies between 14.9 and 17.5 MeV and have widths between 0.2 and 0.3 MeV. The analysis of the measured spectra is hampered by the fact that there is a finite probability for atomic electrons to be kicked out in the reaction. This leads to satellite peaks 88 keV or 15 keV above nuclear excited states with intensities of about 0.1% and 5% for K- and L-electrons respectively. Although the first excited state is at 2.6 MeV it is claimed that all states up to 6.2 MeV of excitation, 150 in total, have been firmly identified and characterized with spin and parity.

**Fission of hyperdeformed states in actinide nuclei**

The successful experimental program at the Q3D spectrograph to study the multiple-humped potential energy landscape in actinide nuclei, that led to the establishment of the existence of a hyperdeformed (axis ratio 3:1) deep third minimum in $^{236}$U [Csa05], was continued. Employing (d,p) reactions in coincidence with fission such hyperdeformed bands were found as well in $^{232}$U [Csi09] and also in the odd-odd nuclei $^{232}$Pa [Csi12] (see Fig. 2) and $^{238}$Np [Csi15]. Absolutely essential for these identifications was the excellent resolution of about 6 keV for the detected protons. Based on the observed level densities the excitation energy of the third potential minimum could be extracted.

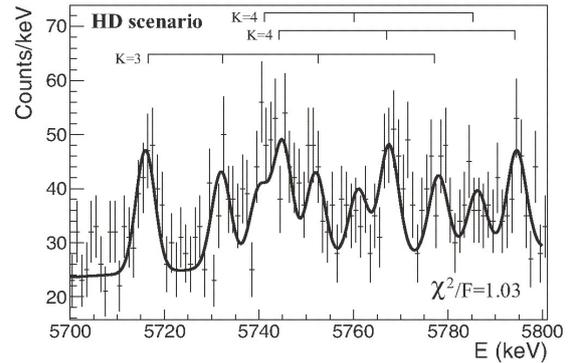

**Figure 2:** *Fission coincident proton spectrum for excitation energies in $^{232}$Pa. Fully drawn is a fit assuming three rotational bands.*

**Clustering in light nuclei**

Clustering of α-particles plays an important role in light nuclei. The prime example is the 1st excited $0^+$ state in $^{12}$C that was postulated by F. Hoyle 60 years ago close to the α-emission threshold and thus responsible for the He burning in stars and the production of carbon. This state is believed to be well represented by an almost linear chain of three α-particles. Clusters consisting of α-particles and neutrons exist as well. In a little heavier systems also molecular structures play a role as pictured in Fig. 3 for the structure of $^{18}$O. We are investigating such systems with the Q3D magnetic spectrograph. Using the ($^7$Li,p) reaction on $^{12}$C and $^{13}$C targets we studied states in $^{18}$O and $^{19}$O, where an α-particle and two neutrons are added to the target nucleus. [Oer10a,Oer10b]. Even far above the particle emission thresholds narrow states exist. Many states are observed for the first time and some can be grouped into rotational bands. With a radioactive $^{14}$C target even cluster states in $^{20}$O were investigated [Boh11]. α-transfer with the ($^6$Li,d) reaction was used to study states in $^{13}$C [Whe12] and $^{16}$O [Whe11]. Here we used in coincidence with deuterons, detected in the focal plane detector of the Q3D, breakup particles of the residual nuclei to measure absolute values of the partial decay widths. The break-up particles were detected in large area, position sensitive Si detectors. Thus, the different breakup channels and even the states in the nucleus, breaking up, can be distinguished. From this information and the particle decay widths some of the observed states can be characterized with respect to their underlying molecular structure. Using the $^9$Be($^3$He,t) reaction excited states in $^9$B have been studied recently [Whe15], which all decay into p+α+α. With sophisticated conditions on the break-up particles the energy and width of the first excited state could be determined indicating to have a shell-model structure rather than a cluster structure.

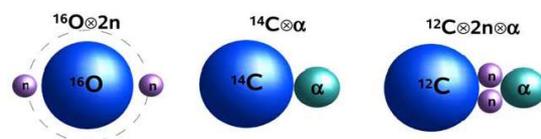

**Fig. 3 :** *Possible molecular configurations of $^{18}$O.*

## Nuclear Astrophysics

Nova explosions occur, when a white dwarf in a binary system has accreted hydrogen from its companion star until a thermonuclear runaway has been triggered with subsequent ejection of material. Thus light elements are formed and dispersed into the interstellar medium. Since the relevant nuclides lie close to the valley of stability, the quantities necessary to calculate the reaction rates as a function of temperature are accessible to high resolution experiments. The calculated element formation can then be compared with the results of infrared and radio observations.

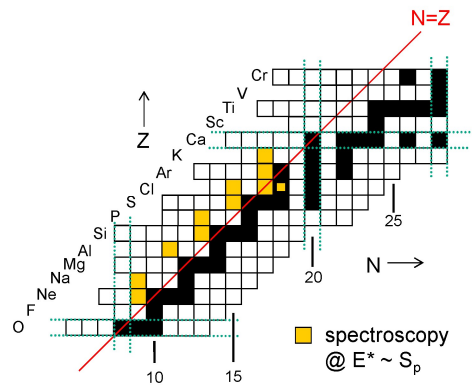

**Figure 3.** *Nuclidic chart: indicated are the nuclei studied with respect to Nova-nucleosynthesis.*

We have used the MLL tandem accelerator together with the Q3D magnetic spectrograph to investigate levels close to the proton emission threshold in nuclei that are formed in the reaction pathway of Novae. Nuclear states in $^{19}$Ne [Lai13, Par15], $^{20}$Na, $^{24}$Al, $^{28}$P, $^{32}$Cl, $^{36}$K [Wre10], $^{27}$Si [Par11b], $^{31}$S [Par11a, Irv13], and $^{34}$Cl [Par09], $^{35}$Cl [Gil17] and $^{35}$Ar [Fry15,Fry17] have been excited with $^3$He and d induced reactions. In many cases targets were used where the target element of interest was implanted into carbon foils at the Univ. of Washington. Precise excitation energies and spin-parity assignments could be obtained even for not resolved states by measuring angular distributions.

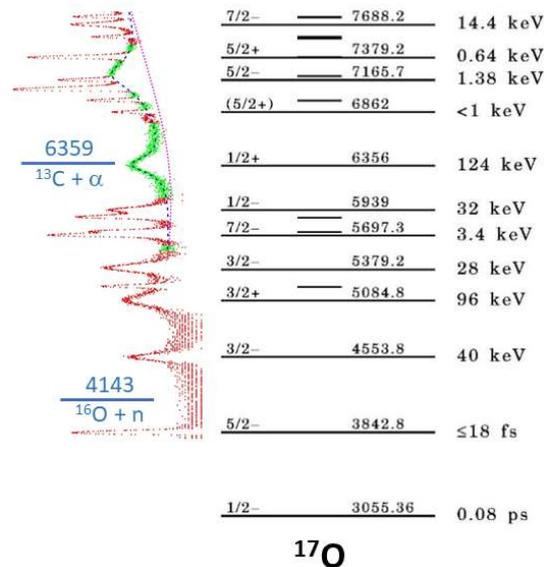

One of the main neutron sources for the s-process is the $^{13}$C($\alpha$,n)$^{16}$O reaction. Thus, states in $^{17}$O close to the $\alpha$-threshold influence the n-production. And states above the n-threshold may act as n-poison through the $^{16}$O(n,$\gamma$)$^{17}$O reaction. We have used the $^{19}$F($\alpha$,d)$^{17}$O reaction to measure the excitation spectrum of $^{17}$O between 3.8 and 7.8 MeV (Fig. 4) and could determine position and total width of the relevant levels with much better precision than before [Fae15], especially for the 1/2$^+$ state just above the $\alpha$-threshold with E*=6363.4(3.1) keV, $\Gamma$=136(5) keV. Also states in $^{21}$Ne in the Gamow-window, which again can act as a neutron-poison, have been investigated using the $^{20}$Ne(d,p)$^{21}$Ne reaction [Nsa16].

**Figure 4.** *Level scheme of $^{17}$O (ENSDF) with the $\alpha$/n-thresholds and the measured spectrum (2 field settings). Shown in green the fit-range for the 6363keV 1/2$^+$ state.*

# Nuclear Structure for the CKM Matrix

The eigenstates of the quarks in weak interaction are not identical to the mass eigenstates. Therefore a unitary transformation is introduced, the CKM matrix. The most precise determination of the first element of the quark mixing matrix comes from nuclear beta-transitions between isobaric analogue states (IAS) with spin 0, where Gamow-Teller transitions are forbidden. Besides the main ingredients, partial half-life and decay energy, also detailed nuclear structure information is necessary for the small, but important corrections. Before the powerful Penning traps came online, precise measurements of reaction Q-values provided the decay energy. A number of values where determined with ($^3$He,t) reactions at the Munich Q3D [Von77]. But it turned out that they were not compatible with later Penning trap values. To test the accuracy of the reaction Q-values, we have determined that for the $^{46}$V decay to $^{46}$Ti [Fae09] using the superb energy resolution of the MLL Q3D magnetic spectrograph. We eliminated most systematic uncertainties by measuring the energy of the $^{46}$Ti($^3$He,t)$^{46}$V reaction relative to $^{47}$Ti($^3$He,t)$^{47}$V* where the final nucleus is also in the excited IAS. The difference in beta-decay Q-values is very small and measured to be 28.73±0.16 keV. A comparison of the recent measurements with the old ones is shown in Fig. 5. The MLL uncertainty (0.27 keV) is dominated by that of the neutron separation energy of $^{47}$Ti (0.3 keV from (n,γ) and 0.25 keV from (d,p) at the MLL). The recent measurements are compatible, whereas the 1977 Munich result is off by more than 3σ. Apparently, the trap magicians [Ero11] constantly improve their precision and it would take a lot of effort to compete using reactions. We also measured the Q-values for the more exotic decays of emitters with N = Z-2: $^{20}$Na, $^{24}$Al, $^{28}$P, and $^{32}$Cl by ($^3$He,t) reactions [Wre10], where the achievable precision is not quite as high. We reached about 1 keV uncertainty and improved the previous precision of 3 – 7 keV considerably.

To test the theoretical calculations of the "charge dependent" corrections, that take care of the violation of isospin symmetry, we investigated the $^{64}$Zn(d$_{pol}$,t)$^{63}$Zn reaction [Lea13a] using polarized deuterons. This is the closest reaction on a stable target to investigate pairing properties in the daughter nucleus of the $^{62}$Ga-$^{62}$Zn decay. For a large number of states in the daughter nucleus the spectroscopic factors could be determined and compared with shell model calculations using the same residual interactions as in the calculations of the charge dependent corrections. Discrepancies can now be used to improve the calculations.

Small branchings of the beta-intensity to excited 0$^+$ states have also to be taken into account. With the $^{64}$Zn(p,t) reaction we searched for excited 0$^+$ states in $^{62}$Zn, and found four such states below 5.4 MeV [Lea13b]. However the previous 0$^+_2$ assignment to a state at 2342keV had to be rejected. This knowledge also restricts the calculations of the charge dependent corrections. To locate excited 0$^+$ states in $^{50}$Cr we have studied the $^{52}$Cr(p,t) reaction [Lea16] and assigned spins based on the angular distribution of the outgoing tritons. With this new information the feeding of excited states can be determined using gamma detection.

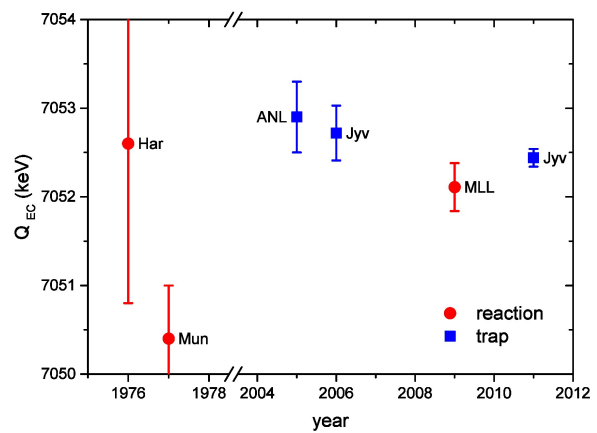

**Figure 5.** *How the precision of the $^{46}$V Q$_{EC}$-value improved with time. The indicated labs are: Harwell, Munich, Argonne National Lab, Jyväskylä and the MLL.*

# Nuclear Structure for Neutrino-less Double Beta-decay

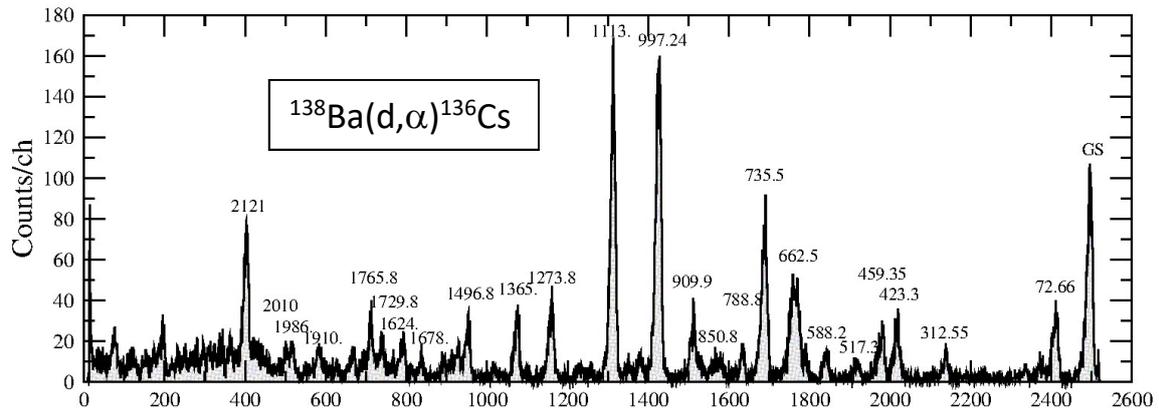

**Figure 6:** *Spectrum of states in $^{136}$Cs.*

If the process of neutrino-less double-β decay (0ν2β) were to be observed, neutrinos would be established as their own antiparticles (Majorana particles) and progress could be made towards determining an absolute scale for the neutrino-mass eigenstates. That neutrinos have mass is established by the observation of neutrino-flavor oscillations. However, such work only establishes differences between the squares of the mass eigenstates. A determination of the lifetime of the 0ν2β decay process would allow access to the absolute mass scale, provided the mechanism responsible for the decay is driven by light Majorana neutrino exchange [Vog12]. The rate of the 0ν2β decay is sensitive to nuclear structure inputs, with the decay rate being proportional to the phase space for the two β-particles, the square of the nuclear transition matrix element and the effective Majorana mass of the electron neutrino. The latter is the sum of the mass eigenstates weighted with the square of the mixing matrix elements. Therefore, from a finite 0ν2β decay rate one gets access to the absolute neutrino mass scale, provided one can reliably calculate the nuclear transition matrix element. As benchmarks to such calculations experimental quantities are necessary.

We have investigated pairing properties for the $^{100}$Mo-$^{100}$Ru system by studying (p,t) reactions on $^{100,102}$Ru and $^{98,100}$Mo targets [Tho12,Free17]. Cross sections were measured at lab angles 6° and 15° to be sensitive to L=0 transitions leading to 0$^+$ states. Whereas the L=0 transfer to and from $^{100}$Ru leads almost exclusively to the ground states, about 20% of the L=0 strength goes to excited 0$^+$ states in the $^{100}$Mo case. Apparently $^{100}$Ru is more on the spherical side of the well-known shape transition in A≈100 nuclei, whereas $^{100}$Mo seems to be more easily deformable. This shape change between the two partners of the 0ν2β decay tends to make the calculation of the matrix element complicated.

Another 0ν2β pair is the $^{150}$Nd-$^{150}$Sm system. For the intermediate nucleus $^{150}$Pm, although next to two stable isobars, there was no knowledge about excited states until 2011. From a ($^3$He,t) and (t,$^3$He) study with moderate energy resolution there were about 20 states known before our own work. We studied excited states in $^{150}$Pm via the $^{152}$Sm(d,α) reaction using our high resolution Q3D spectrograph and with the $^{150}$Nd(p,nγ) reaction at the Bucharest tandem accelerator [Buc12]. We identify in total about 50 levels below 1.5 MeV of excitation.

Similar is the situation for the $^{136}$Xe-$^{136}$Ba system. Until 2011 there were two excited states known in $^{136}$Cs, when a high-energy ($^3$He,t) study with 42 keV resolution yielded 15 excited states up to 3.5 MeV. At the Q3D we studied $^{136}$Cs with the $^{138}$Ba(d,α) reaction up to 2.5 MeV [Reb16] (Fig. 6). Due to the energy loss differences the resolution was only 15 keV, a factor of two more than with a thin target. For the about 25 observed states angular distributions were measured for the assignment of spin and parity.

## Nuclear Structure to Interpret EDM Measurements

The observation of a permanent electric dipole moment (EDM) of a particle would imply a violation of time reversal (T) and parity symmetry (P). The most stringent upper limit to date has been reported for the $^{199}$Hg atom. But to interpret such limits (or possibly finite values) in terms of T-violating couplings of the quarks or their interaction, a detailed knowledge of the nuclear structure is required. To this end we have started to investigate the E1, E2, and E3 strength in $^{199}$Hg and neighbouring nuclei by precision spectroscopy with the Q3D [Gar16]. With the $^{200}$Hg(d,d') reaction the E2 excitation strength of the $2_1^+$ and $4_1^+$ states was measured as well as the E3 strength distribution to a number of $3^-$ states. With the $^{200}$Hg(d,t) pickup reaction we populated states in $^{199}$Hg. We obtained an energy resolution of 7 keV (FWHM) and observed a total of 91 states up to an excitation energy of about 3 MeV, more than half of them were previously not known. From the measured angular distributions we will be able to deduce the transferred angular momentum and the spectroscopic factors for many of these states.

## Nuclear Astrophysics using AMS

For Accelerator-Mass-Spectrometry (AMS) measurements we use two setups. One has a gas-filled 135° dipole magnet (GAMS) for isobar separation followed by an ionization chamber, measuring position, x- and y-angle, total deposited energy and five energy-loss signals [Kni97]. In addition, it is preceded by a time of flight (TOF) measurement. The second installation is a TOF measurement followed by an ionization chamber and a Si.detector for $\delta E$ and $E_{res}$ measurements. From the many applications we shall report only two:

### $^{60}$Fe from Supernovae in the Solar System

Stars with a mass of more than about 8 times the solar mass usually end in a core-collapse supernova explosion (SN). Before and during this explosion new elements, stable and radioactive, are formed by nuclear reactions and a large fraction of their mass is ejected with high velocities into the surrounding space. Most of the new elements are in the mass range until Fe, because there the nuclear binding energies are the largest. If such an explosion happens close to the sun it can be expected that part of the debris might enter the solar system and therefore should leave a signature on the planets and their moons. Already in 1996 we had postulated that $^{60}$Fe ($T_{1/2}$=2.62 Ma [Rug09]) is a promising isotope to search for such SN debris: lighter radioisotopes are abundantly produced by cosmic rays on atmospheric or meteoritic target nuclei up to $^{56}$Fe and $^{60}$Ni, but $^{60}$Fe could only be produced from the rare target isotopes $^{62,64}$Ni. For nuclei with A>80 a considerable background from fission is expected. With our GAMS setup we reach a sensitivity for $^{60}$Fe/Fe concentrations below $10^{-16}$. That is equivalent to detecting the alcohol of a glass of Martini, injected into the lake Constance (after stirring). Indeed, already in 1999 we could report [Kni99] evidence for an increased $^{60}$Fe content in samples of ferromanganese crusts which had grown in the South-Pacific in a depth of 1300 m. We then had analysed well dated crusts from the Central Pacific in depth intervals of 1 – 2 mm, corresponding to age intervals of 0.4 – 0.8 Ma and found a clearly enhanced $^{60}$Fe content between 2 and 3 Ma ago [Kni04]. However, a search in sediments [Fit08], which grow a factor of 1000 faster, did not show a clear enhancement. But recently, an Australian group [Wal16] as well as our lab [Lud16] found an $^{60}$Fe enhancement in sediments from the Indian Ocean and the Pacific for a time period between 1.5 and 3 Ma ago (Fig. 7). This long range indicates that more than one SN have deposited dust in the Solar System. To obtain quantitative information, which is not influenced by atmospheric or oceanic distribution we also analysed Lunar samples [Fim16] which were provided by NASA. Due to the impact of micro-meteorites the Lunar surface is constantly reworked and deposited material is slowly buried. In these samples we also determined the concentration of $^{53}$Mn. Compared to meteoritic samples, where $^{60}$Fe and $^{53}$Mn are produced by cosmic rays in a constant ratio, we measure in most Lunar samples down to a depth of about 10 cm an enhanced (up to a factor of 10) $^{60}$Fe concentration (Fig. 8). This is clear evidence for the $^{60}$Fe input from beyond the Solar system.

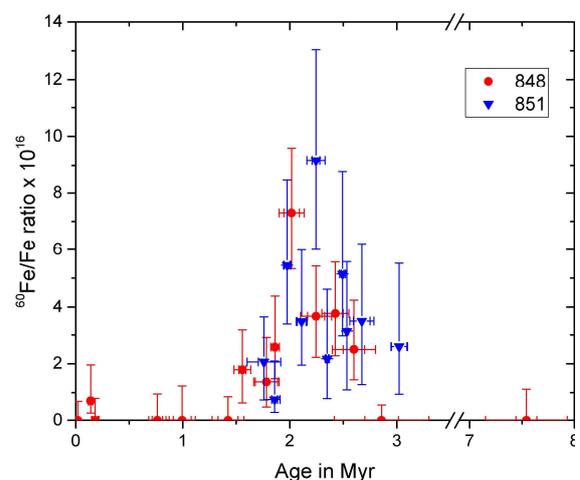

**Figure 7.** *$^{60}$Fe concentration as a function of age for two different sediment cores from the equatorial Pacific [Lud16].*

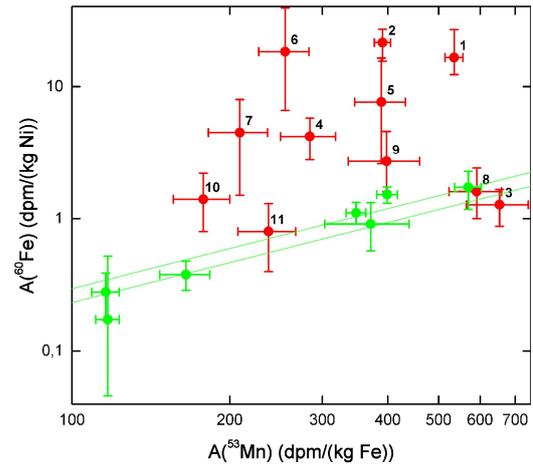

**Figure 8:** $^{60}$Fe and $^{53}$Mn concentrations (relative to their dominant target element for cosmic ray production) for meteoritic (green) and Lunar (red) samples. The green lines indicate the 1σ-band for meteoritic samples.

**Search for primordial heavy elements**

Nuclei heavier than $^{209}$Bi have most often no long-lived isobar. Therefore, it is sufficient to determine the atomic mass for a unique identification. However, ordinary mass spectrometry is not very sensitive because of molecular background. We have searched with AMS at our TOF system for long-lived (primordial) isomers of $^{211-218}$Th, which had been claimed before by a group around A. Marinov, and could disprove their claim. Since Marinov and collaborators had also claimed the discovery of element 112, the attribution of the discovery of this element by the group at GSI was held up. The disproval of the long-lived Th isomers helped the IUPAC/IUPAP joint working party to attribute the discovery and the naming right for the element Copernicium to the GSI group. Another disproval we could accomplish was for primordial $^{244}$Pu, claimed by D.C. Hoffman and her group in 1971 [Hof71]. Since the half-life of $^{244}$Pu is 1/56 of the age of the Solar system, such an observation was surprising. Our result [Lac12] was a 2σ upper limit for the $^{244}$Pu concentration a factor of 6 smaller than the claim by Hoffman. Since Superheavy nuclei close to the magic neutron number 184 could be long-lived and might be reached in the r-process, we also searched for primordial nuclei with mass numbers 292≤A≤310 in possibly homologue elements Os, Pb and "raw-platinum" from a South-African mine. In all cases we reached upper limits for the concentration ratio between $10^{-16}$ and $10^{-14}$.

It should be mentioned here that AMS is also a powerful tool to detect α-emitting actinide nuclei in materials used for large-scale low-background detectors as was done for Cu which is used in the holding structure of the GERDA double-beta experiment [Fam15].

# Materials science

## Test of high density $^{235}$U fuel for the research reactor FRM II

The Munich FRM-II is a HEU-fueled (93 % $^{235}$U) reactor used as a high flux neutron source. Due to the controversy over the use of 93% enriched uranium in the reactor, it was decided to transform to a fuel below 50% $^{235}$U enrichment. The most promising candidates to keep the $^{235}$U density in the fuel about the same but decrease enrichment are molybdenum uranium alloys where the total uranium density is increased. The alloy is either dispersed in an aluminum matrix or as a monolithic foil. TUM researchers uniquely simulate the radiation damage in such alloy systems by irradiating them at a dedicated irradiation setup with 80 MeV $^{127}$I beams up to an ion fluence of $2 \times 10^{17}$ cm$^{-2}$. Although neutron fluences are much larger during such a burn up period it has been shown that the much stronger, non-linear effects related with thermal spike or Coulomb explosion from the high stopping power of the fission fragments challenge the material behavior [Jun13, Wie06]. In particular, interdiffusion layers are formed from which the materials are suffering. Irradiation is performed in $^{235}$U-depleted material models within hours that simulate in pile irradiations of months. In addition, no fission fragments are produced that hinder in-pile irradiated materials from standard materials testing procedures due to its heavy radiation load before months of waiting times.

## High resolution depth profiling of light elements

Our Q3D magnetic spectrograph (Fig. 1) is not only an excellent tool for nuclear physics research it has also unique capabilities in materials analysis using elastic recoil detection (ERD). Hereby, heavy ion beams like 170 MeV $^{127}$I or 40 MeV $^{197}$Au ions are guided under flat angles onto a thin film material to be analyzed. Recoil ions from Rutherford scattering events are analyzed by the Q3D at a scattering angle of 15° or by ΔE-E or TOF-E detectors placed at 40° in the scattering chamber of the Q3D with respect to the incident beam direction. The ERD-technique is in particular well suited to quantitatively analyze profiles of light elements in thin, µm thick films. While with standard detectors a depth resolution of about 10 nm is achieved, the Q3D allows sub-nanometer, in special cases even monolayer, depth resolution [Dol98]. The applications are widespread in materials analysis: ultra-hard and optical materials, ultrathin semiconductor materials or even soft organic matter is analyzed. For example, it was used to determine the elemental composition of ultra-hard nanocomposites leading to a detailed model how the ultra-hardness is developed in such materials [Vep00]. Another example is the analysis of unexpected bimodal range distributions of low energy carbon ions when depositing tetragonal amorphous carbon layers [Neu10]. A very special application of ERD was the determination of the $^{10}$Be/$^{9}$Be ratio in a dilution series, that was a main component in order to accurately measure the half-life of $^{10}$Be to 1.388(0.018) Ma, a common chronometer isotope to determine the age of materials formed on the surface of the earth [Kor10].

## The ion microprobe SNAKE

One of the main instruments used at the Munich tandem accelerator facility is the ion microprobe SNAKE (Superconducting Nanoprobe for Applied nuclear (Kern-) physics Experiments). It makes use of the high brightness ion beams delivered from the tandem accelerator, from protons to heavy ion beams, and a superconducting multipole lens. Sufficient beam can be focused to sub-micrometer beam spots [Gre17] in order to perform various kinds of materials, radiobiology and medical research. In particular, a multicusp ion source features hydrogen and deuterium beams up to 100 µA to be injected into the tandem resulting in a high beam brightness of 2 µA/(mm²mrad²MeV) at SNAKE.

## Hydrogen microscopy

Hydrogen is one of the main elements or an impurity element in nearly any kind of material determining chemical, mechanical, electrical and optical properties. However, nearly all conventional microscopy techniques are not able to analyze lateral distributions of hydrogen in micrometer dimensions. Elastic proton-proton scattering utilizing 10 – 25 MeV protons at SNAKE overcomes several of these limitations. We have demonstrated a hydrogen detection limit below 0.1 at-ppm (< 7 ppb weight) and a lateral resolution of better than 1 µm when analyzing hydrogen embedded in grain boundaries of polycrystalline diamond [Rei04]. The high proton energies are necessary to analyze thin films up to a thickness of a few hundred micrometer since both scattered protons from elastic proton-proton-scattering events are detected in transmission geometry.

The method has been applied to study thin, hydrogen loaded niobium films in view of a model hydrogen storage material deposited on 60 µm thick silicon wafers. After loading the niobium films to an average atomic hydrogen concentration larger than 13 at% the film starts to lift off in the substrate in µm sized regions due to strong local stress while it remains stuck at others. The hydrogen content in the lifted areas increased to 40 at% while it remained at 13 at% on the parts of the film still fixed to the silicon (Fig. 9) [Wag13]. The measurement demonstrates the local separation of the two phases with different hydrogen content. It supports the so called Gorsky effect: due to local stress relaxation, hydrogen diffuses in the resulting gradients of chemical potentials forming the inhomogeneous hydrogen distributions.

Another application is the analysis of minerals from the inner earth mantle. Although they contain hydrogen below 100 at-ppm the integrated amount of hydrogen bound in the inner earth mantle via the minerals are much larger than the hydrogen bound in the oceans. The exact determination of hydrogen contents in these minerals is crucial to understand the mechanical and chemical behavior of the minerals under high pressure and temperature. Proton-proton scattering quantifies hydrogen contents even in small crystals ejected by volcanic activity independent of the analyzed matrix and is used to calibrate standard minerals for commonly used infrared spectroscopy methods or SIMS analysis [Tho15].

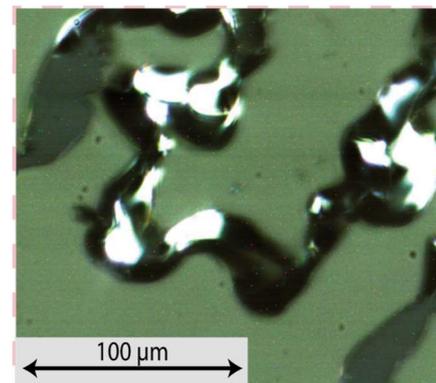
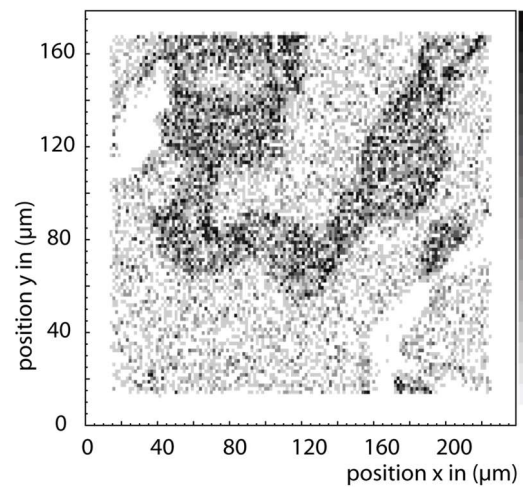

**Figure 9:** *Micrograph of the buckled niobium film (a) analyzed for its local hydrogen content imaged by proton-proton scattering at SNAKE (b). The areas lifted off from the substrate.*

## Radiobiology

The 10-20 MeV protons have a low linear energy transfer in water (LET<10keV) while lithium or carbon ions from SNAKE show high LET. These ions focused to submicrometer diameter are ideally suited to target substructures of cells by counted ions and to directly compare the cellular reactions after low and high LET radiation. Irradiation takes place well defined in space, time, and local dose. Kinetics of cellular processes induced by the ions can be studied immediately after the irradiation event. SNAKE allows to tackle long lasting questions in radiobiology and medicine:

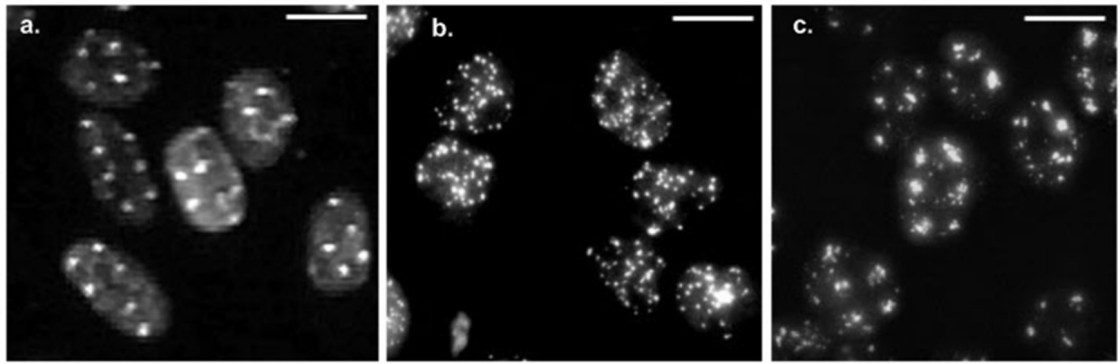

**Figure 10:** *Repair foci (γ-H2AX) show various irradiation modalities to be compared for their effect in creation of chromosomal aberations and also for cell survival. All irradiation modalities were chosen to deposit a mean dose of 1.7 Gy: A grid wise irradiaton of 55 MeV gold ions ions (a), a random irradiation using 20 MeV (low LET) protons (b) and a gridwise irradiation of 117 protons per point depositing the same amount of energy as a single 55 MeV ion (c).*

- **What enhances the radiobiological effectiveness (RBE) for ions with high linear energy transfer (LET)?**

    Alpha particles from radioactive sources or heavy ion beams require less dose to induce the same cellular reactions as low LET reference radiation, for instance X-ray. The ratio of the reference dose $D_{ref}$ to induce the same effect as the dose $D_{LET}$ of high LET radiation defines its RBE: RBE=$D_{ref}$/$D_{LET}$. In order to vary dose distribution on micrometer scale we measured RBE enhancement for induction of dicentric chromosome aberrations of 20 MeV protons (LET=2.6 keV/µm) by focusing a defined number of protons (e.g. 117 per point) to sub-micrometer beam spots in a grid pattern [Sch17]. Thus, the dose distribution on the micrometer scale was artificially changed without changing the mean dose and the dose distribution on the nanometer scale (Fig. 10). The data showed substantial enhancement of dicentrics induction for the grid-wise applied proton spots demonstrating the importance of interaction of DNA double strand breaks (DSB) on the micrometer scale. The data could be in part described by the Monte Carlo code PARTRAC that calculates dicentrics yields using physical dose distribution patterns and models for DNA damage and DNA damage interaction. Cell survival shows similar reduction by focused proton application that could be well fitted using the LEM IV model [Fri18]. The measurements show that the nm, µm and the sizes of cell nuclei of about 10 µm have to be taken into account to model the effects correctly.

- **Microscopic analysis of structure, recruitment kinetics and dynamics of irradiation induced DNA repair foci**
-     When inducing DSB most of them are well repaired by attracting repair factors accumulating in irradiation induced foci. The recruitment kinetics changes for some of these factors in dependence of the ion´s LET while others do not change much [Hab12]. There recruitment and also the movement of these foci can be followed by live cell online fluorescence microscopy at SNAKE when the repair factors are tagged by fluorescent dyes like GFP (green fluorescent protein). The dynamics in the movement of repair foci shows a subdiffusional $t^{0.25}$ behaviour for the distances between foci [Gir13]. This leads to a high probability for the rejoining of the ends of a created DSB but also to join the wrong ends if two DSB are situated within such a small distance.

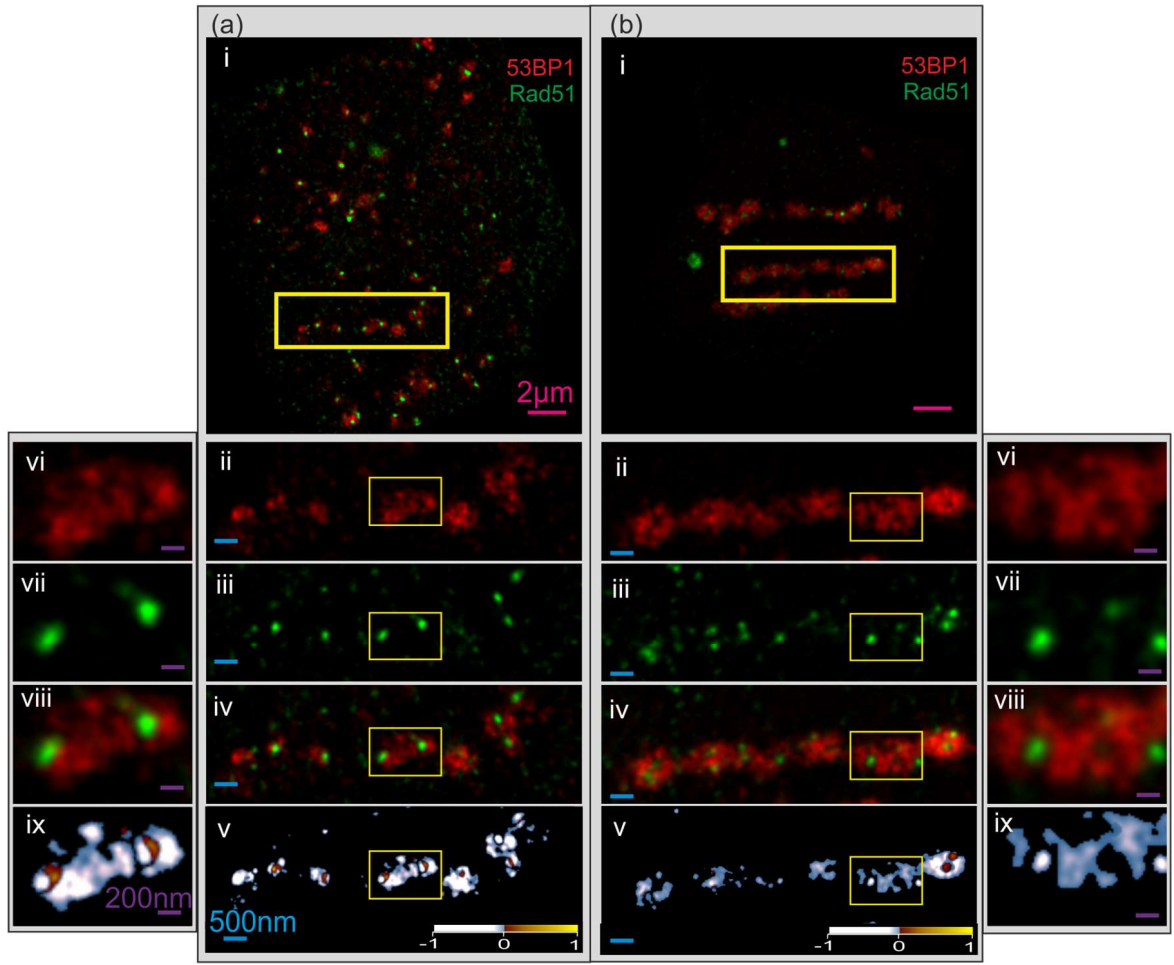

**Figure 11:** *Fine structure of the repair proteins 53BP1 (red) and Rad51 (green) imaged by STED fluorescence microscopy with a lateral resolution of ~ 100 nm. The two foci colocalize on gross scale (i) for proton irradiation (a) as well as for carbon ions. 2 ions traversed the imaged cell of (b).The enlarged images show details of the fine structures for each protein (ii, iii, vi, vii), a merged graph (iv, viii) and a correlation map (v, ix). The correlation map shows mainly anticorrelation in the fine structure of the two proteins showing their exclusion during the repair process [Reindl17].*

The structure of the radiation induced foci is studied using immunofluorescence and ultra high resolution fluorescence STED (STimulated Emission Depletion) microscopy showing fine structures of the different repair factors. Some of the repair factors anti-correlate although the foci colocalize on a gross scale [Rei17]. In addition, the fine structure of the repair foci was analyzed for their relative position to the chromatin structure showing there attraction at the rim of nanometer sized chromatin domains (Fig. 11).

Targeted microbeam irradiation at SNAKE is used to manipulate cellular substructures. We showed, that the irradiation of mitochondria, the power stations of the cell, by high doses led to complete depolarization and inactivation visualized by the darkening of these mitochondria tagged with a polarization sensitive stain (Fig. D4) [Wal17].

# Medical research

**Minibeam proton therapy**

Proton minibeam therapy is investigated for its potential to reduce side effects below that of conventional proton therapy (Fig. 12). High energy proton beams are focused or collimated to submillimeter beam sizes and displaced by distances much larger than the beam sizes leaving tissue between the minibeams with nearly no dose. The beam spreads on the way into the tumor where the beams overlap similar as in conventional pencil beam scanning techniques. Thus, a homogeneous dose deposition is obtained in the tumor even if applying the minibeams from one side (Fig. 12) [Sam17]. Interlacing beams may also be used when using proton minibeams with even larger beam to beam separation from opposing sides such that tissue sparing effects in the healthy tissue can be maintained up to close to the tumor.

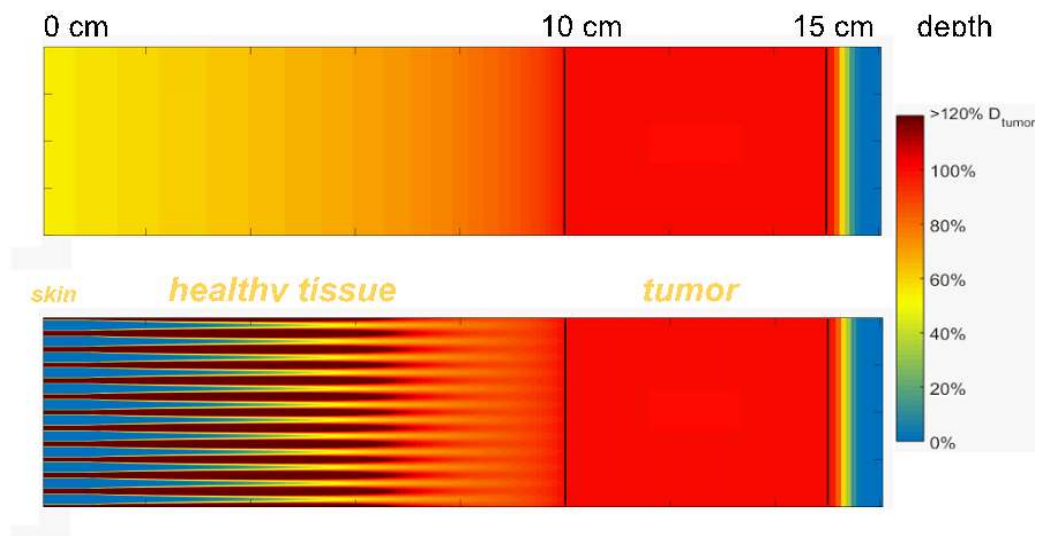

**Figure 12:** *Dose distribution calculated for an extended Bragg peak to treat a hypothetic tumor situated 10cm to 15 cm underneath the skin in case of conventional broad proton beam (upper graph) and proton minibeam (lower graph)*

Inflammation processes where compared after irradiating ears of living mice either by proton minibeams or proton beams of same mean dose of 60 Gy. The minibeams where formed rectangular $0.18 \times 0.18$ mm² at SNAKE and placing the minibeams with inter-beam distances of 1.8 mm. While the broad beam irradiations showed strong inflammation reactions accompanied by strong ear swelling none of these reactions was obtained after proton minibeam irradiation (Fig. 13). When comparing with broad X-ray irradiation of different dose the mean 60 Gy minibeam irradiation was less effective in induction of ear swelling than 10 Gy X-ray irradiation. Thus minibeam irradiation opens new fields to further reduce side effects in healthy tissue but keeping tumor control.

The group of G.D. (UniBW) will install a 70 MeV post-accelerator to produce proton minibeams for further preclinical experiments performed at an image guided small animal irradiation platform in collaboration with K. Parodi (LMU) being supported by an ERC grant.

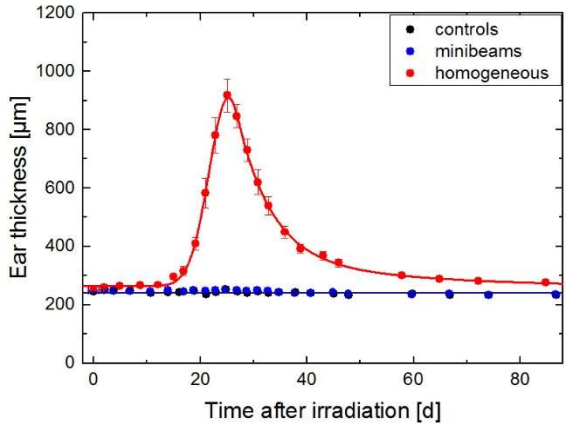

**Figure 13:** *Ear thickness of the right ear of bulb-c mice versus time after irradiation. The mean dose of the irradiated mouse ears was 60 Gy each.*

**In-situ proton range determination in proton therapy**

Although proton and heavy ion beam induces much less ionisations in the healthy tissue compared to X-ray its full potential to spare healthy tissue cannot be used yet since range uncertainties in treatment planning requires large savety margins that destroy somehow the advantageous dose distribution. An elegant way to determine the range of protons online is the use of ionoacustics analyzing the travelling time of ultrasound signals that are generated mainly in the Bragg peak of microsecond pulsed proton beams (Fig. 14). Better than 0.1 mm resolution of the Bragg peak position was demonstrated with 20 MeV proton bunches [Ass15]. Even 3D imaging of the Bragg peak position is possible [Kel16]. An attractive option would be a direct integration into a conventional ultrasonic imaging system to obtain a relative measure of the Bragg peak and tumor position.

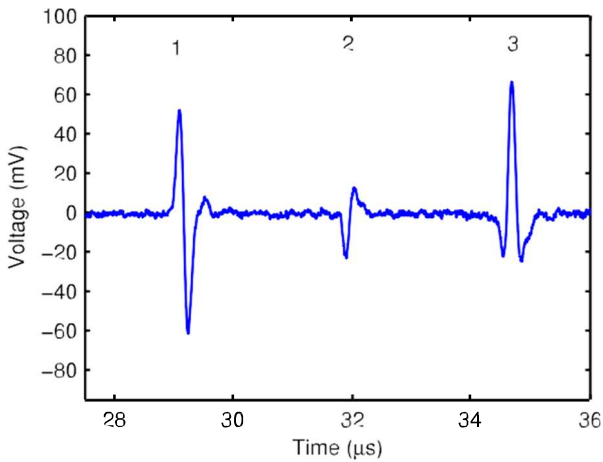

**Figure 14:** *Ionoacoustic signal from 16 ion pulses of 110 ns added from $2 \times 10^6$ protons each transported through a thin kapton window into water. It shows the Bragg peak signal (1), a signal from the entrance window (2) and the Bragg signal reflected from the kapton window traveling the proton range twice [Ass15].*

## Detector Development

The tandem accelerator offers the opportunity to test detectors with beams from protons to $^{238}$U, with intensities of a few Hz up to µA, dc or pulsed with 1 ns resolution, and with well focused or larger area beams broadened in a well localized manner using kHz wobbling. Even secondary neutrons are used.

Direct proton beams allow for studies of spatial detector resolution, ageing behaviour and radiation hardness of detectors or readout components as it is needed in satellite based experiments. Alternatively, well localized proton or deuteron beams can be used to simulate background scenarios as in modern collider experiments allowing hereby for adjustment of the background rate and in a certain range also for the adjustment of the amount of energy

deposited. The availability of wobbled beams with frequencies above 100 Hz enables thus the localized

**for the LHC @ CERN)**

The detectors in the experiments at the LHC are subject to increasing fluxes of particles which demand an enormous rate capability. For the TPC of the heavy ion experiment ALICE a GEM (gas electron multiplier) readout is planned. In order to find the best suited detector gas as well as safe settings of the high voltage, a 20 MeV proton beam, pulsed with 50 kHz and about $10^4$ protons per pulse was used. For the high-luminosity upgrade of ATLAS large tracking detectors on the basis of micromegas are foreseen. These were also tested with up to 550 kHz 20 MeV protons. Similarly, the ATLAS drift tubes have to be made fit for the expected high background rates. Aging tests have been performed with proton beams of up to 100 nA, which deposit in a single night as much charge as ATLAS will see in 10 years of running. No aging was observed. High energy neutrons were produced by breakup of 20 MeV deuterons on Be targets. Neutron fluxes of $10^7/(cm^2s)$ were achieved with an energy distribution around 11 MeV with a width of 8 MeV FWHM.

**for heavy ion machines SIS18@GSI, FAIR@GSI and RIBF@RIKEN**

Although highly segmented silicon detectors nowadays are standard tools in charged particle detection in both, nuclear and particle physics, new readout systems using highest rates at a minimum of power consumption are constantly improved. Full system tests using an easy and fast variation of rates and energy deposit are an important ingredient to reach best performance under the expected extreme conditions which cannot be achieved using standard calibration sources. Examples are the pion tracker [Wir16] developed for the HADES experiment operating at particle rates up to $10^8$ /s, the implantation detectors SIMBA [Hin12] and WAS3ABi [Cel16] developed for high resolution decay spectroscopy experiments at GSI(Darmstadt) and RIKEN (Japan), or the TREX [Bil12] detector telescopes for the Miniball experiment at CERN.

Based on an experiment where the $^{12}C(p,p')^{12}C^*$ reaction was employed to study material performance of large CsI(Tl) crystals, a fundamentally new application of this detector material was found. Light curves not only depend on the doping concentration in the scintillator but also on the ionization density of charged particle tracks. This effect used for many years for particle identification was found to be extremely sensitive to separate punching through from stopped particles. This property called intrinsic phoswich effect [Ben13] allows for two independent measures (total energy deposit and specific energy loss) to characterize particle hits over a wide range of energies and finally led to the design [Cort14] of the multi-million Euro large scale calorimeter CALIFA [Cort14] for the R3B experiment currently set up at FAIR.

**for CRESST (Cryogenic Rare Event Search with Superconducting Thermometers)**

The search for hypothetical dark matter particles requires detectors that can distinguish signals from neutral particles scattered off the nuclei in the detector from signals due to ionizing radiation. Therefore, nearly all dark matter detectors measure for each event two independent signals. In the case of the CRESST detectors not only the phonon signal is detected, but also scintillation light in the $CaWO_4$ crystals, which is less for recoil nuclei than for electrons or photons with the same energy deposit. In addition, there is an increasing amount of quenching of the light output the heavier the target nucleus is off which the neutral particle had scattered.

To calibrate this light output relative to the deposited energy, mono-energetic neutrons are used. These are produced at our Tandem accelerator with the $^1H(^{11}Be,n)^{11}C$ reaction in inverse kinematics. The incident $^{11}B$ energy is chosen such that a resonance in $^{12}C$ at 20.62 MeV is populated which can decay by neutron emission to the $^{11}C$ ground state but just not to the first excited state at 2.000 MeV. Thus mono-energetic neutrons, boosted by the center of mass velocity to 11.6 MeV in 0° direction, are created. These are used for a precise determination of the quenching factors [Jag05,Str14]. The neutrons are scattered off the kryogenic detector and their energy loss is measured by their time of flight using a pulsed beam as the start signal and 40 liquid scintillator detectors for the stop signal. Now a similar dark matter experiment, COSINUS [Ang16], which is based on NaI detectors for phonons and light is also calibrating the quenching factors of their detectors with this neutron source. It served also to characterize pulse shape discrimination and proton quenching in organic liquid scintillators for large scale neutrino detectors [Zim13,Zim15].

**Outlook**

In the past decade colleagues from more than 20 institutions world-wide have collaborated with us on nuclear physics experiments at the Q3D alone. When our lab was closed in 2015 for half a year because of fire protection issues, a distinguished colleague from the Argonne National Laboratory, who has a really profound understanding of nuclear physics and who initiated the programme to determine single-particle occupancies in double-beta decay partners, commented [Sch15]: "I guess it seemed irrational some 40-50 years ago that there were suddenly so many tandems and cyclotrons built for nuclear structure around the world. But it is even more irrational now to see the world heading for literally NO capability in this field."

We gratefully acknowledge discussions and help from R. Gernhäuser, R. Hertenberger, and G. Korschinek. The idea to write this Laboratory Portrait is due to Sissy Körner, the editor of Nuclear Physics News, where this text had to be shortened substantially.